\documentclass[reprint,aip,jcp]{revtex4-1}

\usepackage{hyperref}
\usepackage{amsmath}
\usepackage{amsfonts}
\usepackage{amssymb}
\usepackage{xspace}
\usepackage{graphicx}
\usepackage{multirow}
\usepackage{mathtools}
\usepackage{url}
\usepackage{setspace}
\usepackage{xparse}
\usepackage{subcaption}
\usepackage{dcolumn}
\usepackage{tabularx}
\usepackage[]{todonotes}
\newcolumntype{.}{D{.}{.}{-1}}

\usepackage{tikz}



\usetikzlibrary{positioning}
\usetikzlibrary{calc}
\usetikzlibrary{decorations.pathmorphing}

\DeclareMathAlphabet{\mathpzc}{OT1}{pzc}{m}{it}
\DeclareMathOperator{\erfc}{erfc}
\DeclareMathOperator{\ext}{ext}

\newcommand{\bigOhN}[1]{\ensuremath{\mathcal{O}(N^{#1})}}

\newcommand{\gbk}[2]{\ensuremath{\left( #1 \middle| #2 \right)}}

\newcommand{\si}{\ensuremath{\sigma}\xspace}
\newcommand{\la}{\ensuremath{\lambda}\xspace}

\newcommand{\snbk}[1]{\ensuremath{\left|\gbk{#1}{#1}\right|^{\frac{1}{2}}}\xspace}
\newcommand{\gb}[1]{\ensuremath{\left({#1}\right|}\xspace}
\newcommand{\gk}[1]{\ensuremath{\left|{#1}\right)}\xspace}

\newcommand{\qqrref}{\cite{Maurer:2012p144107}\xspace}
\newcommand{\thws}{\ensuremath{\vartheta_{\mathrm{ws}}}\xspace}
\newcommand{\thsq}{\ensuremath{\vartheta_{\mathrm{SQ}}}\xspace}
\newcommand{\lP}[2]{\ensuremath{P_{#1}^{#2}}\xspace}
\newcommand{\figref}[1]{Figure~\ref{fig:#1}\xspace}
\newcommand{\sqvl}{SQV$\ell$\xspace}
\newcommand{\svl}{SV$\ell$\xspace}
\newcommand{\qvl}{QV$\ell$\xspace}
\newcommand{\Eqref}[1]{Eq.~\eqref{eq:#1}\xspace}
\newcommand{\shS}{\ensuremath{\mathcal S}\xspace}

\begin{document}

\title{A tight distance-dependent estimator for screening three-center Coulomb integrals over Gaussian basis functions}
\author{David~S.~Hollman}
\affiliation{Center for Computational Quantum Chemistry, University of Georgia, 1004 Cedar St., Athens, Georgia 30602, USA}
\affiliation{Department of Chemistry, Virginia Tech, Blacksburg, Virginia 24061, USA}
\author{Henry~F.~Schaefer}
\affiliation{Center for Computational Quantum Chemistry, University of Georgia, 1004 Cedar St., Athens, Georgia 30602, USA}
\author{Edward~F.~Valeev}
\affiliation{Department of Chemistry, Virginia Tech, Blacksburg, Virginia 24061, USA}

\begin{abstract}
  A new estimator for three-center two-particle Coulomb integrals
  is presented. Our estimator is exact for some classes of integrals and is much more efficient than
the standard Schwartz counterpart due to the proper account of distance decay. Although it is not a rigorous upper bound, the maximum degree of underestimation can be controlled by two adjustable parameters. We also give numerical evidence of the excellent tightness of the estimator. The use of the estimator will lead to increased efficiency in reduced-scaling one- and many-body electronic structure theories. 

\end{abstract}

\maketitle

\section{Introduction}

Efficient evaluation of matrix elements of the Coulomb operator (electron repulsion integrals, ERIs) is an essential ingredient of practical quantum chemistry, especially in the context of one-body methods [Hartree--Fock (HF)\cite{Hartree:1947ia} and hybrid Kohn-Sham density functional theory (DFT)\cite{Becke:1993p1372}] and lower-order many-body methods [second-order M{\o}ller--Plesset (MP2)\cite{Moller:1934p618}].
The ERI tensor is defined as
\begin{align}
  g_{AB} \equiv \gbk AB = \int \Omega_A(\vec r_1)\frac1{r_{12}}\Omega_B(\vec r_2)d\tau_1d\tau_2,
  \label{eq:g}
\end{align}
where $\Omega_A$ and $\Omega_B$ are the (Mulliken notation) bra and ket charge distributions.  The most common version of the ERI tensor is the four-center case, in which both $\Omega_A$ and $\Omega_B$ are product densities, composed of either atom-centered basis functions in HF/DFT or molecular orbitals in correlated methods such as MP2.
To reduce the computational cost of the four-center ERIs, $\Omega_A$ and $\Omega_B$ can be approximately expanded in terms of auxiliary basis set; the resulting density-fitting (DF) (also known as the resolution-of-the-identity, RI) approximation
involves also three- and two-center ERIs in which either or both of the bra and ket densities are single, atom-centered functions.

When spatially localized basis functions are used, the number of ERIs with magnitude greater than some threshold $\epsilon$
is \bigOhN2 rather than \bigOhN4, where $N$ is the size of the basis set (and thus proportional to the size of the molecule).
To take advantage of this sparsity it is necessary to estimate the magnitude of ERI very rapidly (i.e.,~significantly faster than it would take to compute it rigorously).
The simplest such estimate invokes the Schwarz inequality\cite{Dyczmons:1973p307,Haser:1989p104,Gill:1994p65,Strout:1995p8448,Schutz:1999p719}
\begin{align}
  \left|\gbk AB\right| \leq \snbk A \snbk B,
  \label{eq:schwarz}
\end{align} 
which holds because \Eqref{g} defines an inner product.

While the Schwarz bound substantially reduces the cost of computing the ERI tensor by screening out small $\Omega_A$ and $\Omega_B$, it is not a tight bound because it neglects the decay over distance between $\Omega_A$ and $\Omega_B$.
Extensive research has been done recently on integral estimates that incorporate the distance between the bra $\gb A$ and the ket $\gk B$ distributions.\cite{Lambrecht:2005p184101,Lambrecht:2005p184102,Doser:2008p3335,Maurer:2012p144107,Maurer:2013p014101} Earlier approaches were based on the multipole expansion of the Coulomb kernel,\cite{Lambrecht:2005p184101,Lambrecht:2005p184102,Doser:2008p3335} but these efforts were later abandoned in favor of an empirical modification of the Schwarz estimate, referred to as the QQR estimator.\cite{Maurer:2012p144107,Maurer:2013p014101}  For four-center ERIs, the QQR estimator can be written as
\begin{align}
  \left|\gbk{\mu\nu}{\la\si}\right| \approx \frac{\snbk{\mu\nu}\snbk{\la\si}}{R-\ext_{\mu\nu}-\ext_{\la\si}}
  \label{eq:qqr}
\end{align}
where $\ext_{\mu\nu}$ and $\ext_{\la\si}$ are the well-separatedness (WS) extents from the continuous fast multipole method (CFMM),\cite{White:1994p8} and $R$ is the distance between centers of charge of bra and ket.

Although the QQR estimator is not an upper bound, it is a tight estimate. In practice, the tightness is more important than the upper bound property anyway because the precision of the computed property (such as energy) is a complex function of the truncation threshold for the
ERIs. Whereas some have investigated how to compute, for example, the Fock matrix with guaranteed precision in a given Gaussian basis,\cite{Rubensson:2008hb,Rudberg:2009ca} guaranteeing precision of the energy and other properties is even more difficult due to their nonlinear dependence on the Hamiltonian via the density matrix/wave function, among other factors. In practice the relationship between precision of the target property and the ERI truncation threshold is established empirically. Underestimation by the ERI estimator will cause some integrals that normally would have been deemed important enough (above the threshold) to be skipped. However, this will only affect the small integrals near the threshold. Furthermore this can be accounted for by adjusting the empirical relationship between precision and truncation threshold. Thus, the lack of the upper-bound property is not an issue in practice. More discussion of the advantages and disadvantages of rigorous upper bounds versus reliably tight bounds can be found in Ref.~\citenum{Maurer:2012p144107}.

In the context of the MP2 method it is necessary to replace the ``effective'' distance in the denominator of the QQR estimate
with its second or third power to take into account the vanishing leading-order multipoles
of bra and ket densities.\cite{Maurer:2013p014101}
The same issue arises when we want to estimate the magnitude of the three-center integrals in the context of
the context of reduced-scaling electronic structure methods that utilize (local) density fitting approximations.\cite{Kussmann:2013iy,Bowler:2012p036503,Riplinger:2013p034106,Hansen:2011ek,Neese:2009db,Neese:2009p114108,Hollman:2013p064107,Sodt:2006p194109,Adler:2009p054106,Adler:2009p241101,Hattig:2012p204105,Tew:2011p074107,Usvyat:2013hr,Adler:2011p144117,Izmaylov:2008p3421,Aquilante:2009p154107} Indeed, \Eqref{qqr} is not a good estimator for the two- and three-electron integrals over atom-centered Gaussian basis functions
because one or more of the lowest-order multipoles of $\Omega_A$ and/or $\Omega_B$ will vanish for many integrals, and hence long-range decay with $R$ will be significantly faster.

To take advantage of the sparsity of the three-center integrals, we developed a rapid estimator with proper asymptotic $R$-dependence. 
Although our estimator differs technically from the QQR estimator of Ochsenfeld and co-workers,\qqrref it is similar in spirit. The primary objective of the estimator is tightness, not a rigorous upper bound. However, the underestimates in our new bound can be reduced arbitrarily by adjusting the WS parameter.  The estimator is also governed by a second parameter, $\thsq$, which loosely controls the trade-off between tightness of the estimator and the maximum amount of underestimation.

\section{Theory}

In this work, we will deal with the basis of atom-centered contracted real solid harmonic Gaussian-type orbitals (GTOs), defined as\cite{HelgakerBook,Rico:2012cf}
\begin{align}
  &\ \ \ \ \ \chi_\mu(\vec r; \vec R) =  \sum_{a} c_{a,\mu} \phi_{a\ell_\mu m_\mu}(\vec r - \vec R), \\
  &\ \ \ \ \ \ \ \  \phi_{a}(\vec r) = \shS_{\ell_a m_a}(\vec r) \exp(-\zeta_a |\vec r|^2) \\
  &\begin{multlined}
  \shS_{\ell  m}(r, \vartheta, \varphi) =
  (-1)^{ m}N_{\ell m}(\zeta_a)\\
  \qquad\ \ \ \ \ \ \ \ \ \  \times r^\ell \lP\ell{| m|}(\cos\vartheta)\begin{dcases}
    \cos( m\varphi)  &  m \geq 0 \\
    \sin (|m|\varphi)  &  m < 0
  \end{dcases}
  \end{multlined} 
  \label{eq:sharm}
  \\
  &\ \ \ \ \ N_{\ell m}(\zeta_a) = \left[\frac{\pi\Gamma(\ell+\frac32)(\ell+|m|)!}{(2\zeta_a)^{\ell+\frac32}(2\ell + 1)(\ell-|m|)!}(1+\delta_{m0})\right]^{-\frac12}
  \label{eq:nfact}
\end{align}
where $c_{a,\mu}$ are the contraction coefficients, $\lP\ell{m}$ are the associated Legendre polynomials, and $\Gamma(\ell+\frac32)$
is the gamma function [since $\ell$ is a nonnegative integer, $\Gamma(\ell+\frac32) = \sqrt{\pi} (2\ell+1)!!/2^{\ell+1}$].

\subsection{Notation}

Throughout this work we will use Mulliken (chemists') bra-ket notation shown in Eq.~\eqref{eq:g}.  Three-center ERI will be written with the principal (orbital) basis product $\gb{\mu\nu}$ in the bra and the one-center auxiliary (density fitting) basis function $\gk X$ in the ket. Greek letters $\mu$ and $\nu$ will be used to denote functions in the principal basis, and $X$ will be used to denote basis functions in the auxiliary basis.  For primitives, we will use $a$ and $b$ for the principal basis and $c$ for the auxiliary basis.

\subsection{The Primitive $\gbk{ss}X$ Case}
\label{sec:svl}

Consider a three-center ERI $(ab|c)$, where primitive GTOs $\phi_a$, $\phi_b$, and $\phi_c$ have principal angular momentum quantum numbers $\ell_a=0$, $\ell_b=0$, and arbitrary $\ell_c$.  We use the notation $m_a$, $m_b$, and $m_c$ for the $L_z$ quantum numbers; $\zeta_a$, $\zeta_b$, and $\zeta_c$ for the primitive exponents; and $\vec R_a$, $\vec R_b$, and $\vec R_c$ for the centers of the respective primitives (analogous notation will be used throughout).  According to the Gaussian product theorem (GPT), the bra $(ab|$ can be rewritten as a single Gaussian centered at the center-of-exponent point $\vec R_p$ located on the line in between $\vec R_a$ and $\vec R_b$ (temporarily neglecting normalization):
\begin{align}
  \phi_a(\vec r_a)\phi_b(\vec r_b) &= \exp(-\zeta_a|\vec r_a|^2)\exp(-\zeta_b|\vec r_b|^2) \\
  &= K_{ab} \exp(-\zeta_p|\vec r_p|^2) \\
  &\equiv \Omega_A(\vec r_p),
\end{align}
where $\zeta_p = \zeta_a + \zeta_b$, $\vec R_p = (\zeta_a\vec R_a +\zeta_b \vec R_b)/\zeta_p$, $\vec r_i \equiv \vec r - \vec R_i$, and $K_{ab} \equiv e^{-\zeta_{a}\zeta_{b} |\vec R_{a} - \vec R_{b}|^{2}/\zeta_{p}}$.  Since $\Omega_A$ is spherically symmetric, from a large enough distance away it can be viewed as a point charge centered at $\vec R_p$ of magnitude $S_{ab}$ (overlap between $\phi_a$ and $\phi_b$).

Now consider the potential generated at $\vec R_p$ by a primitive Gaussian $\phi_c$ located at $\vec R_{c}$.  For practical purposes, we wish to determine the average potential of an entire shell (i.e., all $m_c$ with $-\ell_c \leq m_c \leq \ell_c$) rather than an individual $m_c$.  Thus, without loss of generality we can orient our system with the $z$ axis along the vector $\vec R \equiv \vec R_{p} - \vec R_{c}$. With this orientation, a nonzero potential is generated only by the $m_{c}=0$ Gaussian, and we are left with (omitting normalization of $\phi_{c}$ for now)
\begin{align}
\tilde V_{\ell_c}(R) = \int \frac{r^\ell}{|\vec r - \vec R|}\exp(-\zeta_c r^2) P_{\ell_c}(\cos \vartheta )d\vec r.
  \label{eq:pot1}
\end{align}
We then use the multipole (Laplace) expansion of the Coulomb operator
\begin{align}
  \frac 1{|\vec r - \vec R|} = \sum_{k=0}^\infty \frac{r_{<}^k}{r_{>}^{k+1}}P_k(\cos\vartheta),
\end{align}
where $\vartheta$ is the angle between $\vec r$ and $\vec R$, $r_< = \min(R,r)$, and $r_> = \max(R,r)$,
to obtain the potential:
\begin{align}
\tilde V_{\ell_c}(R) &= \begin{multlined}[t]
  \frac{4\pi}{2\ell + 1}\Bigg[\int_0^R\frac{r^{2\ell+2}}{R^{\ell+1}}\exp(-\zeta_c r^2)dr
  \\
  + \int_R^\infty R^\ell\exp(-\zeta_c r^2)dr\Bigg]
\end{multlined} \\
&= \begin{multlined}[t]
  \frac{4\pi}{2\ell + 1}\bigg[\frac{\Gamma\!\left(\ell_c+\frac32\right)-\Gamma\!\left(\ell_c+\frac32, R^2\zeta_c\right)}{2R^{\ell+1}\zeta^{\ell+\frac32}} 
  \\ + \frac{\exp(-\zeta_c R^2)R^{\ell_c}}{2\zeta_c}\bigg],
  \end{multlined}
  \label{eq:pot2}
\end{align}
where $\Gamma(s,x)$ is the (upper) incomplete gamma function.  The second term in Eq.~\eqref{eq:pot2} rapidly decays to zero for reasonable values of $\zeta_c$.  Since we are interested in behavior at large $R$ (i.e., where the bra can be viewed as a point charge), we can omit this term.  Similarly, the incomplete gamma function $\Gamma(s,x)$ is bounded from above by $\Gamma(s)$ for real, positive $x$, and the second part of the first term in Eq.~\eqref{eq:pot2} rapidly decays to zero with distance $R$ for reasonable values of $\zeta_c$.  Thus, we have
\begin{align}
  \tilde V_{\ell_c}(R) \approx \frac{2\pi\Gamma\!\left(\ell_c + \frac32\right)}{(2\ell_c+1)R^{\ell+1}\zeta_c^{\ell_c+\frac32}}
\end{align}
Inclusion of the normalization factor from Eq.~\eqref{eq:nfact} gives 
\begin{align}
  V_{\ell_c}(R) &= \tilde V_{\ell_c}(R)N_{\ell_c 0}(\zeta_c) \\
  & \approx \left(\frac{2}{\zeta_c}\right)^{\frac{2\ell_c+3}4}
  \sqrt{\frac{2\pi\Gamma\!\left(\ell_c + \frac32\right)}{2\ell_c+1}}
  \ \,R^{-\ell_c-1}
  \\
  &=
  \frac{(2\pi)^{\frac34}\sqrt{(2\ell_c-1)!!}}{\zeta_c^{\frac{2\ell_c+3}4}R^{\ell_c+1}}
  \label{eq:pot3}
\end{align}
To arrive at our long-range estimate for $\gbk{ab}c$, we simply scale Eq.~\eqref{eq:pot3} by the magnitude of the charge term from the bra to get
\begin{align}
  \left|\gbk{ab}c\right| &\approx |S_{ab}| \, V_{\ell_c}(R), &\text{(large $R$)} \\
  &\approx
  |S_{ab}| \, \frac{(2\pi)^{\frac34}\sqrt{(2\ell_c-1)!!}}{\zeta_c^{\frac{2\ell_c+3}4}R^{\ell_c+1}}
  , &\text{($\ell_a=\ell_b=0$)}.
  \label{eq:sexact}
\end{align}
We will refer to this formula as the \svl estimator. We derived it by considering the classical limit of the interaction of the electrostatic potential of the $|X)$ ket with the point-charge representation of the $(ss|$ ket; however, since the leading-order multipole of a product of two arbitrary Gaussians of any angular momenta is also a charge, at a large-enough distance away this estimate should also be sufficiently accurate.

\subsection{Extension to Arbitrary $\ell_a$ and $\ell_b$}
\label{sec:qvl}

Ideally, we would like to replace the charge-like contribution from the bra in Eq.~\eqref{eq:sexact} with a multipole expansion of the bra charge distribution to handle the general case.  Unfortunately, it is too expensive to expand the bra distribution to multipoles of high enough order simply for the purpose of estimating integrals.  Almost a decade of research on this topic\cite{Lambrecht:2005p184101,Ochsenfeld:2007vw,Kussmann:2007p204103,Beer:2008p221102,Maurer:2012p144107,Maurer:2013p014101,Kussmann:2013iy} has led Ochsenfeld and co-workers to the conclusion that the best substitute for the multipole expansion in this context is the Schwarz bound,
\begin{align}
  Q_{ab} = \snbk{ab}
\end{align}
(and hence the QQR estimator in Eq.~\eqref{eq:qqr}).\cite{Maurer:2012p144107}  In the $(ss|$ case, the Schwarz estimate can be directly related to the overlap by
\begin{align}
  S_{ab} = Q_{ab} \left(\frac{\pi}{2 (\zeta_a+\zeta_b) }\right)^{\frac14}.
\end{align}
For higher angular momenta $\ell_a$ and $\ell_b$, though, $Q_{ab}$ incorporates contributions from higher-order multipoles, thus improving the estimate (as discussed in ref.~\citenum{Maurer:2012p144107}).  Incorporating this prefactor, we arrive at the estimator
\begin{align}
  \left|\gbk{ab}{c}\right|&\approx Q_{ab}\, \frac{\pi\sqrt{2(2\ell_c-1)!!}}{R^{\ell_c+1}\zeta_c^{\frac{2\ell_c+3}{4}}(\zeta_a + \zeta_b)^{\frac14}}, & \text{(large $R$)}
\end{align}
which we call the \qvl estimator.  Again, we wish to estimate entire shells for practical purposes, so we have taken shell-wise Frobenius norms to obtain a $Q_{ab}$ that is rotationally invariant:
\begin{align}
  Q_{ab} = \sqrt{\sum_{m_a = -\ell_a}^{\ell_a}\sum_{m_b = -\ell_b}^{\ell_b} (Q_{a(m_a),b(m_b)})^2},
\end{align}
where $a(m_a)$ and $b(m_b)$ are the functions with $L_z$ quantum numbers $m_a$ and $m_b$ in the {\it shells} with indices $a$ and $b$.  Note that this enforces rotational invariance since the Frobenius norm is invariant under rotations.  It can easily be shown that when $\vec R_a \ne \vec R_b$, the GPT product distribution will always have an $\ell = 0$ contribution.  Thus, for large enough $R$, the charge-like term of the bra multipole expansion will dominate, and the integral will eventually decay as $R^{-\ell_c+1}$, since the denominators of higher-order multipole terms will become much larger than this term.

\subsection{Defining ``Large $R$''}

Up to this point, we have been vague about the definition of ``large $R$,'' stating only that it is a distance from which a bra of the form $(ss|$ may be approximated as a point charge to sufficient precision.  To a first approximation, this concept is already well established in the context of CFMM.\cite{White:1994p8}  In CFMM, two distributions $\Omega_{ab}$ and $\Omega_{cd}$ are considered ``well-separated'' if 
\begin{align}
  R_{ab,cd} > \ext_{ab} + \ext_{cd}
\end{align}
where the extents $\ext_{ab}$ and $\ext_{cd}$ are given by
\begin{align}
  \ext_{ab} = \sqrt{\frac{2}{\zeta_a+\zeta_b}}\, \erfc^{-1}(\thws)
\end{align}
for some given WS threshold $\thws$.  While this formula is obtained for spherical Gaussians, this definition turns out to be sufficient for our purposes, given the other approximations involved in our QV$\ell$ estimator.  For distributions that are not well separated, the QV$\ell$ estimator reverts to Schwarz screening (Eq.~\eqref{eq:schwarz}).  Thus, the QV$\ell$ estimate can be summarized as
\begin{align}
  \left|\gbk{ab}{c}\right|&\approx \begin{dcases}
    Q_{ab}\, \frac{\pi\sqrt{2(2\ell_c-1)!!}}{R^{\ell_c+1}\zeta_c^{\frac{2\ell_c+3}{4}}(\zeta_a + \zeta_b)^{\frac14}}, & \!\!R > \ext_{ab} + \ext_{c} \\
    Q_{ab}Q_c, & \!\!R \leq \ext_{ab} + \ext_{c}
  \end{dcases}
\end{align}

\subsection{Combining $S_{ab}$ and $Q_{ab}$}
\label{sec:sqvl}
As demonstrated in Section~\ref{sec:results}, the \qvl estimator is robust, yielding estimates within a factor of 10 or so of the exact value for the vast majority of three-center ERIs.  However, there are several aspects of the formulation that are fundamentally dissatisfying.  While $Q_{ab}$ has replaced $S_{ab}$ to better account for the effects of higher order multipoles, the distance scaling factor remains that of the zeroth order multipole---namely, the overlap.  But the reasoning behind this was that the term with slowest decay  should dominate for large enough $R$.  Hence $S_{ab}$ should be a better representation of the bra contribution than $Q_{ab}$ at large-enough separations. The problem is that when $a$ and $b$ are close together but differ significantly in angular momentum, the threshold beyond which higher order multipoles are negligible is much larger than $\ext_{ab} + \ext_{c}$.  The CFMM extents only indicate when it is safe to approximate each term in the multipole expansion of the integral by point multipole interactions, not where it is necessarily safe to truncate the multipole expansion at the leading-order term.  When the separation is such that higher order multipoles {\em are} actually small enough compared to the overlap, the \svl estimator is much better than the \qvl estimator, since it gives the proper prefactor to the proper term.  Incorporating this concept into the extents for the purposes of thresholding would require the computation of higher order multipole integrals, which we have already noted is too expensive for our purposes.  Instead, one can roughly determine the importance of higher order multipoles by taking the ratio $S_{ab}/Q_{ab}$.  When this ratio is small, higher-order multipole effects will overcome the additional $R$ factors in the denominators of the hypothetical multipole expansion, warning us that the exclusion of the higher-order contributions from $Q_{ab}$ could be dangerous.  However, if this ratio is large enough, the dominant contribution to the multipole expansion will be the overlap, and $S_{ab}$ should be used to approximate the bra contribution.  We conclude that the estimator should be controlled by an additional screening parameter, $\thsq$.  Defining a common prefactor for notational convenience:
\begin{align}
  \beta_{\ell}(\zeta) \equiv \zeta^{-\frac{2\ell+3}4}\sqrt{(2\ell-1)!!},
\end{align}
we can now introduce our best estimator for three-center ERIs in terms of performance and flexibility, which we will call the \sqvl estimator:
\begin{align}
  \left|\gbk{ab}{c}\right|\approx 
  \begin{dcases}
    |S_{ab}|\frac{(2\pi)^{3/4}\,\beta_{\ell_c}\!(\zeta_c)}{R^{\ell_c+1}}
    & \!\!\begin{multlined}[c]
      R > \ext_{ab} + \ext_{c}  \\ \text{and } S_{ab}/Q_{ab} > \thsq 
    \end{multlined} 
    \\
    Q_{ab}\frac{\pi\sqrt2\,\beta_{\ell_c}\!(\zeta_c)}{(\zeta_a+\zeta_b)^{\frac14}R^{\ell_c+1}}
    & \!\!\begin{multlined}[c]
      R > \ext_{ab} + \ext_{c} \\ \text{and } S_{ab}/Q_{ab} \leq \thsq 
    \end{multlined} 
    \\
    \ Q_{ab}Q_c & \!\! R \leq \ext_{ab} + \ext_{c}.
  \end{dcases}
  \label{eq:sqvl}
\end{align}
In other words, the \sqvl estimator ``interpolates'' between the \svl and \qvl estimators:
in the limits $\thsq\rightarrow0$ and $\thsq\rightarrow\infty$ the \sqvl estimator becomes equivalent to the \svl and \qvl estimators, respectively.

As an aside, we note that for the purposes of our discussion here, three-center ERIs do not include cases where $\vec R_\mu = \vec R_\nu$ coincidentally.  While these integrals are indeed part of the full three-center ERI tensor, they are a small enough part that the discussion of these integrals can be neglected in the current context.  The \sqvl estimator gives an approximate bound for these integrals, but because of the angular momentum addition rules, the actual decay with distance is sometimes much more rapid than the estimate accounts for, leading to significant overestimation.  A better estimate for the special two-center case could be developed, but from a practical standpoint it is not worth the effort.

\subsection{Contracted Basis Functions}
\label{sec:contr}
Thus far, our discussion has focused only on primitive basis functions.  For practical purposes, the extension to contracted basis functions is unimportant for many basis sets, since in many cases contracted basis functions are used to represent core orbitals, which do not contribute significantly to long-range integrals.  Nevertheless, the extension of the \sqvl estimator to contracted basis functions is relatively trivial, and yet it performs reasonably well even for basis sets composed entirely of contracted functions (see Section~\ref{sec:results}).  Following Ochsenfeld, et al.,\cite{Maurer:2012p144107} we typically define contracted extents $\ext_{\mu\nu}$ as
\begin{align}
  \ext_{\mu\nu} = \max_{a\in\mu, b\in\nu}\left\{\ext_{ab} +\, r_{ab,\mu\nu}\right\},
\end{align}
where $r_{ab,\mu\nu}$ is the distance from the GPT center of the primitive pair $|ab)$ and the coefficient weighted center of charge of the product $|\mu\nu)$.  However, in the case of generally contracted basis sets such as the ano-pV$X$Z series,\cite{Neese:2011el} this formulation will substantially overestimate most of the extents, and a more careful (but less safe) formula is needed.  For these basis sets, we use a coefficient-weighted average of the primitive pair extents:
\begin{align}
  \ext_{\mu\nu} = \frac{\sum\limits_{a\in\mu, b\in\nu}c_{a,\mu}c_{b,\nu}(\ext_{ab} + r_{ab,\mu\nu})}{\sum\limits_{a\in\mu, b\in\nu}c_{a,\mu}c_{b,\nu}}
\end{align}
where $c_{a,\mu}$ and $c_{b,\nu}$ are contraction coefficients.  For the purposes of determining $\zeta_a$, $\zeta_b$, and $\zeta_c$ for, e.g.,~Eq.~\eqref{eq:sqvl}, the most diffuse exponent in the contraction is used for both standard and generally contracted basis sets.  It is possible that a more efficient ``effective exponent'' formula could be developed, but a thorough investigation of screening for generally contracted basis sets is beyond the scope of this work.


\section{Computational Details}

The \svl, \qvl, and \sqvl estimators were implemented in a development version of the Massively Parallel Quantum Chemistry ({\tt MPQC})\cite{mpqc} package.  We tested our estimates on a test set of three different molecular systems---benzene tetramer ($\pi$-stacked geometry with an inter-monomer separation of $3.2$\AA), linear icosane, and a cluster of 29 water molecules (Cartesian coordinates in supplemental information)---with five different basis set/auxiliary basis set pairs:  cc-pVDZ\cite{Dunning:1989p1007} with cc-pVTZ/JK,\cite{Weigend:2002p4285} cc-pVTZ with cc-pV5Z/JK, Def2-SVP\cite{Weigend:2005dh} with Def2-SVP/C,\cite{Weigend:1998p143,Hellweg:2007fs} aug-cc-pVTZ with aug-cc-pVTZ-RI,\cite{Weigend:2002jp} and ano-pVDZ with aug-ano-pVTZ.\cite{Neese:2011el}  Since the chemistry of the molecules tested is less relevant to the current context (since the density matrix or other chemically important quantities are not involved in these estimates), the variety of basis sets with a variety of coefficients, exponents, and contraction schemes is more important to assessing the quality of the estimators than the variety of molecules examined.  All error statistics were assessed with respect to shell-wise Frobenius norms for both the estimates and the actual integral values.  All bra shell pairs were prescreened with a Schwarz threshold of $10^{-10}$; that is, a bra pair $\gb{\mu\nu}$ was excluded from all statistics if 
\begin{align}
  \snbk{\mu\nu} < \frac{10^{-10}}{\max\snbk{X}}.
\end{align}
While the choice of this pair prescreening threshold has a small effect on the averages and standard deviations of the statistics, the worst case behaviors are largely unaffected by this choice, since these usually arise from pairs composed of functions with different angular momenta on neighboring atoms.  These pairs usually have relatively large Schwarz estimates and are not affected by pair prescreening.


\section{Results and Discussion}
\label{sec:results}

\subsection{Qualitative Performance}

\def\thepicname{test_set_cc-pVTZ_cc-pV5Z_JKFIT_ws_1e-04_sq_1e-01}

\begin{figure}
  \includegraphics[width=0.4\textwidth]{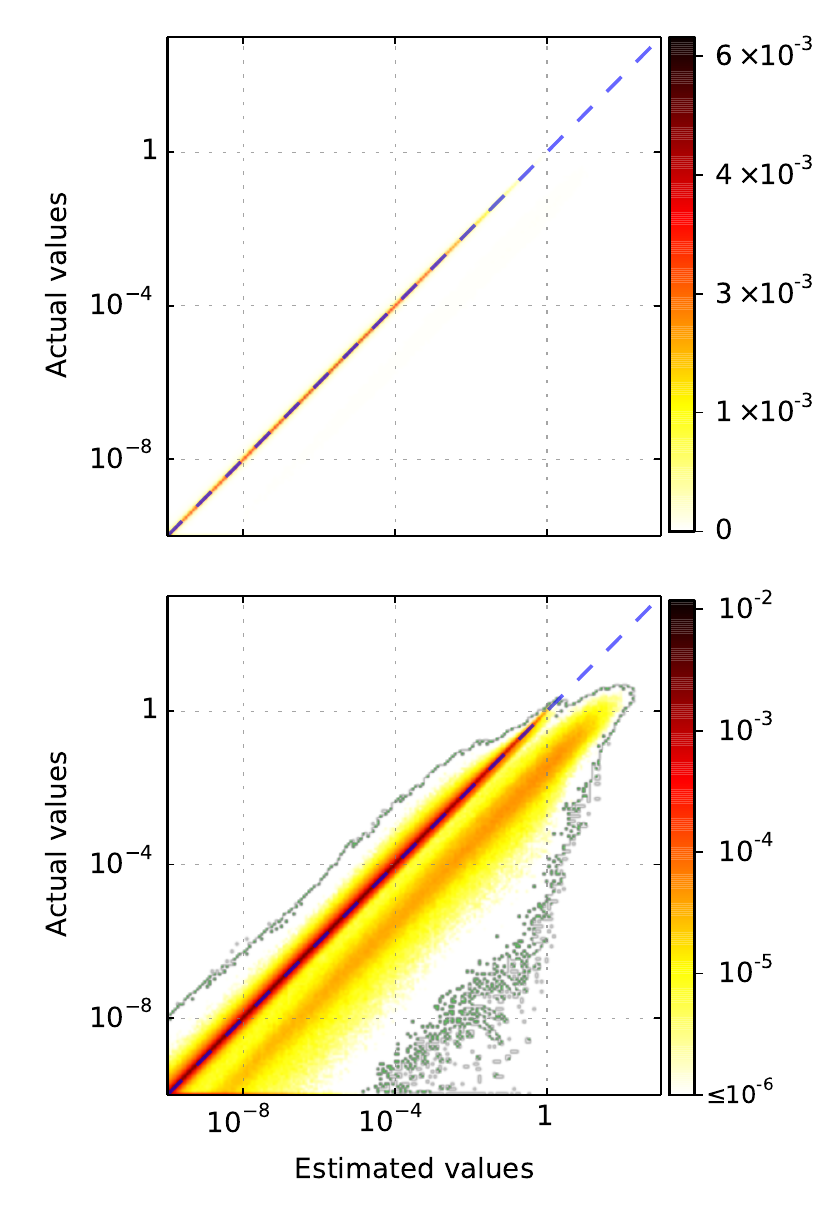}
  \caption{Heat map of \sqvl estimated versus exact integral shell norms obtained for our standard test set of molecules with the cc-pVTZ/cc-pV5Z/JK bases, with $\thws = 10^{-4}$ and $\thsq=10^{-1}$.  The plots use a linear (top) and logarithmic (bottom) color scale, with color values representing the fraction of total integrals in a given 2D histogram bin (of which there are 200 horizontal and 200 vertical).  The green line in the bottom plot shows the boundary between the mostly empty histogram bins and the completely empty bins.
  }
  \label{fig:eva_all}
\end{figure}

\begin{figure*}
  \includegraphics[width=\textwidth]{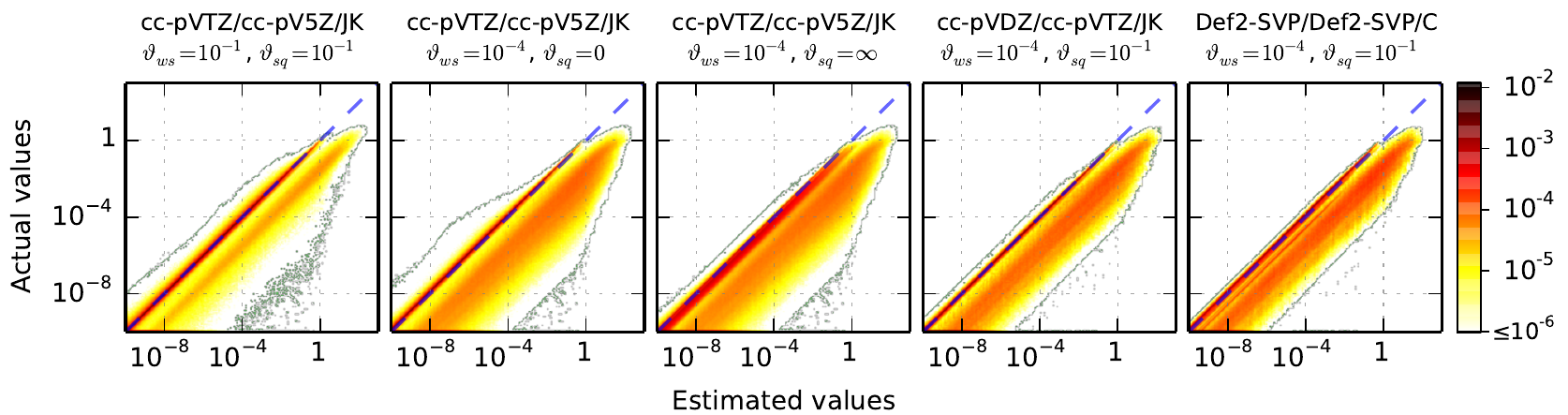}
  \caption{The same plot from \figref{eva_all} but with a different $\thws$, $\thsq$, or basis set.  See caption of \figref{eva_all} for details, and note that the color scale here is logarithmic.  Note also the slightly expanded color scale relative to \figref{eva_all} needed to accommodate the greater variety of basis sets.}
  \label{fig:eva_alt}
\end{figure*}

\figref{eva_all} shows a heat map of the estimated versus exact shell norms obtained with the \sqvl estimator ($\thws = 10^{-4}$, $\thsq = 10^{-1}$). The data were obtained for our standard test set of molecules and the cc-pVTZ/cc-pV5Z/JK basis set pair.  Using a linear color scale (top plot), the estimator appears perfect; i.e., all data points fall along the ideal $I_{\mathrm{estimate}} = I_{\mathrm{actual}}$ line. A more complete picture of performance appears only with a logarithmic color scale (bottom plot). There are two bands in the plot. The first band follows closely the ideal line and contains the vast majority of the data points; it can be identified with the well-separated integrals (cases 1 and 2 in \Eqref{sqvl}). The minor band below the ideal line contains the data points where the estimated values are significantly greater than the exact ones, which is typical of the Schwarz-estimated integrals (case 3 in \Eqref{sqvl}).  Note that the \sqvl estimator eliminates the vast majority of the overestimates that would occur with the pure Schwarz screening.  With still larger molecules or less diffuse basis sets, the Schwarz band would be even less prominent.

Since \sqvl is not an upper bound, a few integrals are {\em under}estimated (the data above the ideal line); however this is an exceedingly rare occurrence. Keep in mind that the cc-pVTZ principal basis set and, particularly, the cc-pV5Z/JK auxiliary basis set are much larger and utilize much higher angular momentum than those typically used for large molecule computations.  Indeed, the underestimated data points are almost not visible on heat maps for cc-pVDZ/cc-pVTZ/JK and Def2-SVP/Def2-SVP/C (fourth and fifth plots, \figref{eva_alt}); i.e., these plots look similar that for a hypothetical rigorous upper bound.

The second and third plots in \figref{eva_alt} show the difference between the pure \svl estimator ($\thsq = 0$) and the pure \qvl estimator ($\thsq = \infty$).  The \qvl shows much less density of integrals in the underestimation region above the main diagonal, but at the expense of a noticeable broadening of the estimates along the main diagonal relative to the \svl or \sqvl estimators.  Finally, we note that in the first plot of \figref{eva_alt}, the increase in the WS parameter relative to the data in \figref{eva_all} does not change the picture much at all.  For the molecules in our test set, the majority of the integrals that are well-separated with $\thws=10^{-1}$ are also well-separated with $\thws=10^{-4}$, so the difference is indistinguishable in this representation of the data. However, the outer limits indicating the worst over- and underestimates are basically identical for both $\thws$ values; this suggests that the worst case behavior can be attributed to a poor representation of higher-order multipoles in the bra rather than a lack of well-separatedness.

\begin{figure*}
  \includegraphics[width=\textwidth]{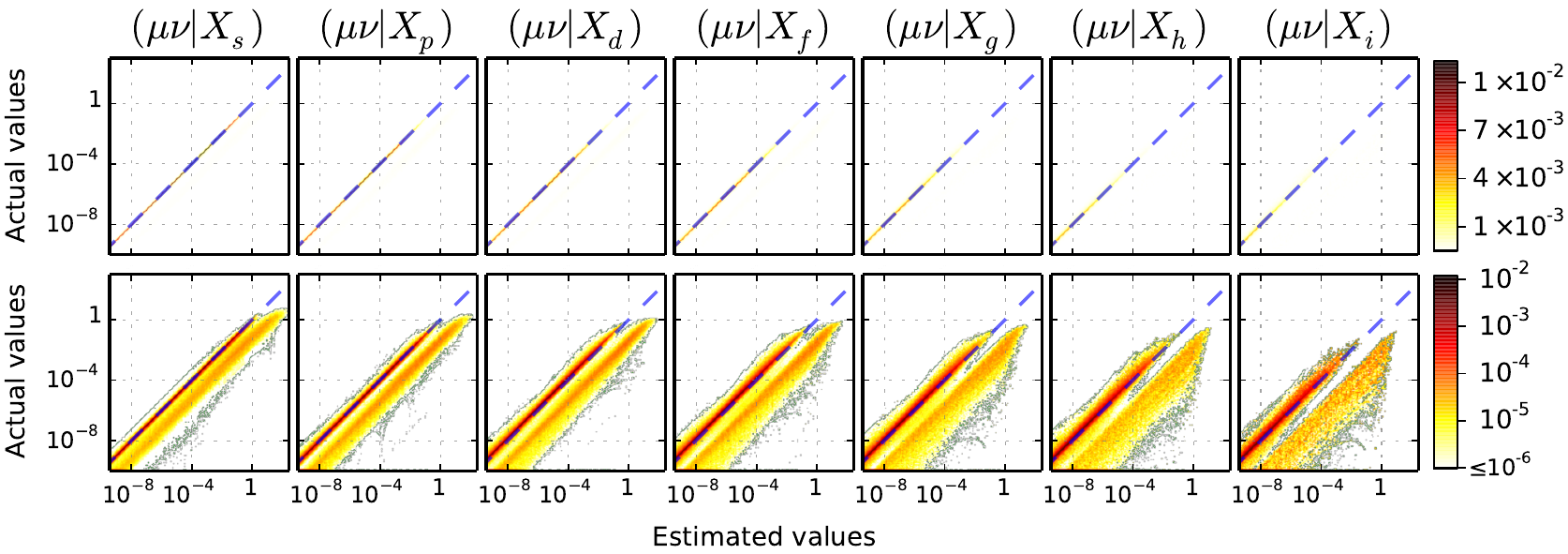}
  \caption{Data from Figure \ref{fig:eva_all} split by the angular momentum $\ell_X$ of the auxiliary basis function $\gk X$.  The color shows the fraction of integrals {\em with a given $\ell_X$} in a particular histogram bin (rather than the fraction of all integrals, as in \figref{eva_all}).  See caption of Figure~\ref{fig:eva_all} for details.}
  \label{fig:eva_am}
\end{figure*}

\figref{eva_am} shows the data from \figref{eva_all} split by angular momentum of the ket $\gk X$.  Notice that the main diagonal broadens with increasing $\ell_X$.  This is attributed to higher-order multipoles from the bra becoming more relevant with respect to their interaction with the ket as $\ell_X$ increases.  In other words, integrals with larger $\ell_X$ are inherently harder to estimate, which is not surprising given that they are {\em significantly} more expensive to compute.  As expected, the Schwarz band broadens and shifts further from the ideal as $\ell_X$ increases, because the distance factor (omitted from the Schwarz estimate) becomes more important to an accurate estimate with increasing $\ell_X$.

\subsection{Quantitative Performance}

Performance of the integral estimators can be measured by analyzing large samples of the ratio of the estimate to the actual integral value: \begin{align}
  F = I_{\mathrm{estimate}}/I_{\mathrm{actual}},
\end{align}
where $I_{\mathrm{estimate}}$ and $I_{\mathrm{actual}}$ are the norms for the estimator and the actual computed integral shells, respectively (Ref. \onlinecite{Maurer:2012p144107} used symbol $F$ to denote this ratio, hence we will follow this notation for consistency). 

\begin{table*}
  \caption{Statistics for the ratio $F = I_{\mathrm{estimate}}/I_{\mathrm{actual}}$ for our test set of three molecules (see text) and various basis sets, $\thws$ values, and $\thsq$ values.  Note that the $\thsq = 0$ case corresponds to the pure SV$\ell$ estimator (see Section~\ref{sec:svl}) and the $\thsq = \infty$ case corresponds to the pure QV$\ell$ estimator (see Section~\ref{sec:qvl}).  Note that, while $N_{ws}$ should remain exactly constant for a given $\thws$, in practice it varies slightly, because cases where the Schwarz estimate is smaller than the \sqvl estimate are not counted as well-separated.}
  \label{tab:err}
\begin{ruledtabular}
    \begin{tabular}{ccccccc.r}
Basis & Aux. Basis & \multicolumn{1}{c}{\ensuremath{\thws}} & \multicolumn{1}{c}{\ensuremath{\thsq}} & \multicolumn{1}{c}{\ensuremath{\bar F}} & \multicolumn{1}{c}{\ensuremath{\sigma(\log F)}} & \multicolumn{1}{c}{\ensuremath{F_{\mathrm{min}}}} & \multicolumn{1}{c}{\ensuremath{F_{\mathrm{max}}}} & \multicolumn{1}{c}{\ensuremath{N_{\mathrm{ws}}/10^6}} \\
\hline
\multirow{8}{*}{cc-pVDZ} & \multirow{8}{*}{cc-pVTZ/JKFIT} & \multicolumn{1}{c}{\ensuremath{10^{-1}}} & \multicolumn{1}{c}{\ensuremath{0}} & 1.004 & 0.078 & 0.036 & 6.726 & 44.1 \\
 &  & \multicolumn{1}{c}{\ensuremath{10^{-1}}} & \multicolumn{1}{c}{\ensuremath{0.1}} & 1.022 & 0.088 & 0.036 & 24.515 & 44.1 \\
 &  & \multicolumn{1}{c}{\ensuremath{10^{-1}}} & \multicolumn{1}{c}{\ensuremath{0.5}} & 1.303 & 0.175 & 0.044 & 24.515 & 44.3 \\
 &  & \multicolumn{1}{c}{\ensuremath{10^{-1}}} & \multicolumn{1}{c}{\ensuremath{\infty}} & 1.542 & 0.166 & 0.050 & 24.515 & 44.6 \\
 &  & \multicolumn{1}{c}{\ensuremath{10^{-4}}} & \multicolumn{1}{c}{\ensuremath{0}} & 1.001 & 0.058 & 0.083 & 3.080 & 31.0 \\
 &  & \multicolumn{1}{c}{\ensuremath{10^{-4}}} & \multicolumn{1}{c}{\ensuremath{0.1}} & 1.023 & 0.074 & 0.199 & 24.485 & 31.0 \\
 &  & \multicolumn{1}{c}{\ensuremath{10^{-4}}} & \multicolumn{1}{c}{\ensuremath{0.5}} & 1.336 & 0.175 & 0.199 & 24.485 & 31.2 \\
 &  & \multicolumn{1}{c}{\ensuremath{10^{-4}}} & \multicolumn{1}{c}{\ensuremath{\infty}} & 1.576 & 0.162 & 0.295 & 24.485 & 31.4 \\
\cline{3-9}
\multirow{8}{*}{cc-pVTZ} & \multirow{8}{*}{cc-pV5Z/JKFIT} & \multicolumn{1}{c}{\ensuremath{10^{-1}}} & \multicolumn{1}{c}{\ensuremath{0}} & 0.995 & 0.117 & \multicolumn{1}{c}{\ensuremath{{2.8} \times 10^{-4}}} & 2293.872 & 177.6 \\
 &  & \multicolumn{1}{c}{\ensuremath{10^{-1}}} & \multicolumn{1}{c}{\ensuremath{0.1}} & 1.143 & 0.142 & 0.002 & \multicolumn{1}{c}{\ensuremath{{3.7} \times 10^{4}}} & 177.6 \\
 &  & \multicolumn{1}{c}{\ensuremath{10^{-1}}} & \multicolumn{1}{c}{\ensuremath{0.5}} & 1.907 & 0.249 & 0.002 & \multicolumn{1}{c}{\ensuremath{{3.7} \times 10^{4}}} & 180.4 \\
 &  & \multicolumn{1}{c}{\ensuremath{10^{-1}}} & \multicolumn{1}{c}{\ensuremath{\infty}} & 2.081 & 0.222 & 0.003 & \multicolumn{1}{c}{\ensuremath{{3.7} \times 10^{4}}} & 181.2 \\
 &  & \multicolumn{1}{c}{\ensuremath{10^{-4}}} & \multicolumn{1}{c}{\ensuremath{0}} & 0.990 & 0.087 & 0.001 & 2293.872 & 129.9 \\
 &  & \multicolumn{1}{c}{\ensuremath{10^{-4}}} & \multicolumn{1}{c}{\ensuremath{0.1}} & 1.175 & 0.130 & 0.037 & \multicolumn{1}{c}{\ensuremath{{3.7} \times 10^{4}}} & 130.0 \\
 &  & \multicolumn{1}{c}{\ensuremath{10^{-4}}} & \multicolumn{1}{c}{\ensuremath{0.5}} & 2.004 & 0.247 & 0.042 & \multicolumn{1}{c}{\ensuremath{{3.7} \times 10^{4}}} & 132.4 \\
 &  & \multicolumn{1}{c}{\ensuremath{10^{-4}}} & \multicolumn{1}{c}{\ensuremath{\infty}} & 2.167 & 0.218 & 0.042 & \multicolumn{1}{c}{\ensuremath{{3.7} \times 10^{4}}} & 133.0 \\
\cline{3-9}
\multirow{8}{*}{Def2-SVP} & \multirow{8}{*}{Def2-SVP/C} & \multicolumn{1}{c}{\ensuremath{10^{-1}}} & \multicolumn{1}{c}{\ensuremath{0}} & 1.459 & 0.211 & 0.103 & 20.615 & 28.8 \\
 &  & \multicolumn{1}{c}{\ensuremath{10^{-1}}} & \multicolumn{1}{c}{\ensuremath{0.1}} & 1.483 & 0.215 & 0.144 & 128.377 & 28.8 \\
 &  & \multicolumn{1}{c}{\ensuremath{10^{-1}}} & \multicolumn{1}{c}{\ensuremath{0.5}} & 1.927 & 0.264 & 0.198 & 128.377 & 28.9 \\
 &  & \multicolumn{1}{c}{\ensuremath{10^{-1}}} & \multicolumn{1}{c}{\ensuremath{\infty}} & 2.196 & 0.259 & 0.198 & 128.377 & 29.1 \\
 &  & \multicolumn{1}{c}{\ensuremath{10^{-4}}} & \multicolumn{1}{c}{\ensuremath{0}} & 1.446 & 0.203 & 0.138 & 12.000 & 19.3 \\
 &  & \multicolumn{1}{c}{\ensuremath{10^{-4}}} & \multicolumn{1}{c}{\ensuremath{0.1}} & 1.478 & 0.208 & 0.390 & 128.377 & 19.3 \\
 &  & \multicolumn{1}{c}{\ensuremath{10^{-4}}} & \multicolumn{1}{c}{\ensuremath{0.5}} & 1.970 & 0.262 & 0.456 & 128.377 & 19.4 \\
 &  & \multicolumn{1}{c}{\ensuremath{10^{-4}}} & \multicolumn{1}{c}{\ensuremath{\infty}} & 2.240 & 0.255 & 0.514 & 128.377 & 19.5 \\
\cline{3-9}
\multirow{8}{*}{aug-cc-pVTZ} & \multirow{8}{*}{aug-cc-pVTZ-RI} & \multicolumn{1}{c}{\ensuremath{10^{-1}}} & \multicolumn{1}{c}{\ensuremath{0}} & 1.002 & 0.159 & \multicolumn{1}{c}{\ensuremath{{7.4} \times 10^{-5}}} & 2293.872 & 532.1 \\
 &  & \multicolumn{1}{c}{\ensuremath{10^{-1}}} & \multicolumn{1}{c}{\ensuremath{0.1}} & 1.846 & 0.245 & 0.011 & \multicolumn{1}{c}{\ensuremath{{4.7} \times 10^{4}}} & 532.8 \\
 &  & \multicolumn{1}{c}{\ensuremath{10^{-1}}} & \multicolumn{1}{c}{\ensuremath{0.5}} & 2.726 & 0.311 & 0.012 & \multicolumn{1}{c}{\ensuremath{{4.7} \times 10^{4}}} & 538.4 \\
 &  & \multicolumn{1}{c}{\ensuremath{10^{-1}}} & \multicolumn{1}{c}{\ensuremath{\infty}} & 2.986 & 0.271 & 0.018 & \multicolumn{1}{c}{\ensuremath{{4.7} \times 10^{4}}} & 540.5 \\
 &  & \multicolumn{1}{c}{\ensuremath{10^{-4}}} & \multicolumn{1}{c}{\ensuremath{0}} & 0.986 & 0.128 & \multicolumn{1}{c}{\ensuremath{{1.8} \times 10^{-4}}} & 2293.872 & 336.8 \\
 &  & \multicolumn{1}{c}{\ensuremath{10^{-4}}} & \multicolumn{1}{c}{\ensuremath{0.1}} & 2.130 & 0.260 & 0.038 & \multicolumn{1}{c}{\ensuremath{{4.7} \times 10^{4}}} & 336.8 \\
 &  & \multicolumn{1}{c}{\ensuremath{10^{-4}}} & \multicolumn{1}{c}{\ensuremath{0.5}} & 3.129 & 0.324 & 0.111 & \multicolumn{1}{c}{\ensuremath{{4.7} \times 10^{4}}} & 341.2 \\
 &  & \multicolumn{1}{c}{\ensuremath{10^{-4}}} & \multicolumn{1}{c}{\ensuremath{\infty}} & 3.368 & 0.282 & 0.122 & \multicolumn{1}{c}{\ensuremath{{4.7} \times 10^{4}}} & 342.6 \\
\cline{3-9}
\multirow{8}{*}{ano-pVDZ} & \multirow{8}{*}{aug-ano-pVTZ} & \multicolumn{1}{c}{\ensuremath{10^{-1}}} & \multicolumn{1}{c}{\ensuremath{0}} & 3.770 & 0.274 & 0.001 & \multicolumn{1}{c}{\ensuremath{{2.4} \times 10^{5}}} & 65.7 \\
 &  & \multicolumn{1}{c}{\ensuremath{10^{-1}}} & \multicolumn{1}{c}{\ensuremath{0.1}} & 3.866 & 0.268 & 0.024 & \multicolumn{1}{c}{\ensuremath{{2.4} \times 10^{5}}} & 65.7 \\
 &  & \multicolumn{1}{c}{\ensuremath{10^{-1}}} & \multicolumn{1}{c}{\ensuremath{0.5}} & 4.659 & 0.292 & 0.036 & \multicolumn{1}{c}{\ensuremath{{3.5} \times 10^{5}}} & 66.6 \\
 &  & \multicolumn{1}{c}{\ensuremath{10^{-1}}} & \multicolumn{1}{c}{\ensuremath{\infty}} & 5.784 & 0.304 & 0.060 & \multicolumn{1}{c}{\ensuremath{{6.1} \times 10^{5}}} & 66.4 \\
 &  & \multicolumn{1}{c}{\ensuremath{10^{-4}}} & \multicolumn{1}{c}{\ensuremath{0}} & 3.639 & 0.267 & 0.001 & \multicolumn{1}{c}{\ensuremath{{6.9} \times 10^{4}}} & 49.6 \\
 &  & \multicolumn{1}{c}{\ensuremath{10^{-4}}} & \multicolumn{1}{c}{\ensuremath{0.1}} & 3.739 & 0.264 & 0.024 & \multicolumn{1}{c}{\ensuremath{{6.9} \times 10^{4}}} & 52.0 \\
 &  & \multicolumn{1}{c}{\ensuremath{10^{-4}}} & \multicolumn{1}{c}{\ensuremath{0.5}} & 4.569 & 0.290 & 0.036 & \multicolumn{1}{c}{\ensuremath{{3.5} \times 10^{5}}} & 49.7 \\
 &  & \multicolumn{1}{c}{\ensuremath{10^{-4}}} & \multicolumn{1}{c}{\ensuremath{\infty}} & 5.696 & 0.298 & 0.060 & \multicolumn{1}{c}{\ensuremath{{3.5} \times 10^{5}}} & 49.5 \\
\end{tabular}
  \end{ruledtabular}

\end{table*}

Table~\ref{tab:err} shows several statistical measures of $F$ obtained from all unique shell triplets
generated from our molecular test set with several different basis sets, and several relevant values of the estimator parameters ($\thws$ and $\thsq$).  First, the estimator usually performs worse for larger basis sets than the smaller ones, as noted earlier.  The exception is the cc-pVDZ/cc-pVTZ/JK pair, which outperforms the Def2-SVP/Def2-SVP/C pair in terms of $\bar F$, $F_{\mathrm max}$, and $\sigma(\log F)$ (though the $F_{\mathrm min}$ values are slightly better for the latter pair).  This exception is attributed to the presence of contracted functions in the Def2-SVP/C basis, while the cc-pVTZ/JK basis is completely uncontracted.  Our simple handling of contracted estimates (as discussed in Section~\ref{sec:contr}) is to blame here.  One could use the estimator on individual primitive triplets and then carry out the contraction, but we feel that this is a significant increase in effort for only a marginal increase in performance.  Indeed, though the ano-pVDZ/aug-ano-pVTZ results are by far the worst, they are impressively tenable given the massive simplification from Section~\ref{sec:contr} (massive in the context of generally contracted basis sets, that is).

\begin{table}[h]
  \caption{Statistics for the ratio $F = I_{\mathrm{estimate}}/I_{\mathrm{actual}}$ for our test set of three molecules with various $\thsq$ parameters and split by angular momentum $\ell_X$ of the auxiliary ket shell $\gk X$, with the cc-pVTZ\cite{Dunning:1989p1007} basis and the cc-pV5Z/JK\cite{Weigend:2002p4285} auxiliary basis and $\thws = 10^{-4}$.}
  \label{tab:erram}
  \begin{ruledtabular}
    \begin{tabular}{cccccrr}
$\thsq$ & \multicolumn{1}{c}{\ensuremath{\ell_X}} & \multicolumn{1}{c}{\ensuremath{\bar F}} & \multicolumn{1}{c}{\ensuremath{\sigma(\log F)}} & \multicolumn{1}{c}{\ensuremath{F_{\mathrm{min}}}} & \multicolumn{1}{c}{$F_{\mathrm{max}}$} & \multicolumn{1}{c}{\ensuremath{N_{\mathrm{ws}}/10^6}} \\
\hline
\multirow{7}{*}{$0$} & 0 & 1.009 & 0.047 & 0.013 & 2293.87 & 43.5 \\
 & 1 & 0.996 & 0.067 & 0.006 & 19.20 & 27.8 \\
 & 2 & 0.987 & 0.087 & 0.004 & 693.32 & 24.1 \\
 & 3 & 0.975 & 0.108 & 0.002 & 28.33 & 17.2 \\
 & 4 & 0.962 & 0.130 & 0.002 & 16.88 & 10.9 \\
 & 5 & 0.943 & 0.158 & 0.001 & 15.72 & 5.3 \\
 & 6 & 0.920 & 0.184 & 0.001 & 9.25 & 1.2 \\
\cline{2-7}
\multirow{7}{*}{$0.1$} & 0 & 1.282 & 0.122 & 0.296 & ${3.7} \times 10^{4}$ & 43.5 \\
 & 1 & 1.176 & 0.122 & 0.145 & 345.38 & 27.8 \\
 & 2 & 1.134 & 0.127 & 0.100 & 693.32 & 24.1 \\
 & 3 & 1.091 & 0.133 & 0.080 & 334.74 & 17.2 \\
 & 4 & 1.057 & 0.144 & 0.062 & 312.09 & 10.9 \\
 & 5 & 1.026 & 0.162 & 0.042 & 269.45 & 5.3 \\
 & 6 & 0.992 & 0.182 & 0.037 & 165.29 & 1.2 \\
\cline{2-7}
\multirow{7}{*}{$0.5$} & 0 & 2.112 & 0.247 & 0.500 & ${3.7} \times 10^{4}$ & 44.1 \\
 & 1 & 2.006 & 0.246 & 0.282 & 345.38 & 28.2 \\
 & 2 & 1.977 & 0.246 & 0.193 & 4299.39 & 24.6 \\
 & 3 & 1.917 & 0.246 & 0.111 & 334.74 & 17.5 \\
 & 4 & 1.877 & 0.248 & 0.071 & 312.09 & 11.2 \\
 & 5 & 1.842 & 0.251 & 0.042 & 269.45 & 5.5 \\
 & 6 & 1.800 & 0.258 & 0.064 & 165.29 & 1.2 \\
\cline{2-7}
\multirow{7}{*}{$\infty$} & 0 & 2.275 & 0.218 & 0.500 & ${3.7} \times 10^{4}$ & 44.3 \\
 & 1 & 2.169 & 0.217 & 0.282 & 345.38 & 28.4 \\
 & 2 & 2.138 & 0.217 & 0.204 & 4299.39 & 24.7 \\
 & 3 & 2.078 & 0.217 & 0.121 & 334.74 & 17.6 \\
 & 4 & 2.038 & 0.218 & 0.071 & 312.09 & 11.3 \\
 & 5 & 2.003 & 0.221 & 0.042 & 269.45 & 5.6 \\
 & 6 & 1.958 & 0.227 & 0.064 & 165.29 & 1.3 \\
\end{tabular}
  \end{ruledtabular}

\end{table}

Also note that while the average and standard deviation improve noticeably with the reduction of the $\thsq$ parameter, the $F_{\mathrm{min}}$ gets substantially worse.  This is a result of the exclusion of higher order effects from the QV$\ell$ estimator which compensate for poor behavior in the worst edge cases at the expense of the average case.  Similarly, the decrease in the $\thws$ parameter comes at the cost of a roughly 30-40\% decrease in the number of well-separated integrals (and thus, in the number of integrals accessible to the distance-dependent part of the estimator).

Table \ref{tab:erram} shows $F$ statistics for the shell triplets of cc-pVTZ/cc-pV5Z/JK grouped according to the ket angular momentum $\ell_X$.  Due to the narrower variety of exponents and contraction schemes for the higher angular momentum in the cc-pV5Z/JK basis set, higher angular momentum estimates are more statistically accurate with respect to mean and $F_{\mathrm{max}}$, though the former is likely also a result of some error cancellation between over- and underestimates (as evidenced by the $\thsq = 0$ and $\thsq = 0.1$ cases).  However, the minimum ratios for higher angular momentum are lower.  This latter behavior is anticipated, since one would expect the importance of higher order multipoles in the bra to be more pronounced for integrals with higher angular momentum in the ket.  Also, as the parameter $\thsq$ increases, so does the consistency of the standard deviation with respect to angular momentum.  For the pure SV$\ell$ case, we see a much broader distribution for $\ell_X=6$ than for $\ell_X=0$, while in the pure QV$\ell$ case, the standard deviation is nearly identical for all angular momenta.  Again, this is attributed to the greater importance of higher-order effects for larger angular momenta.

\subsection{Performance Versus Bra-Ket Distance}

\begin{figure}
  \centering
  \includegraphics[width=0.5\textwidth]{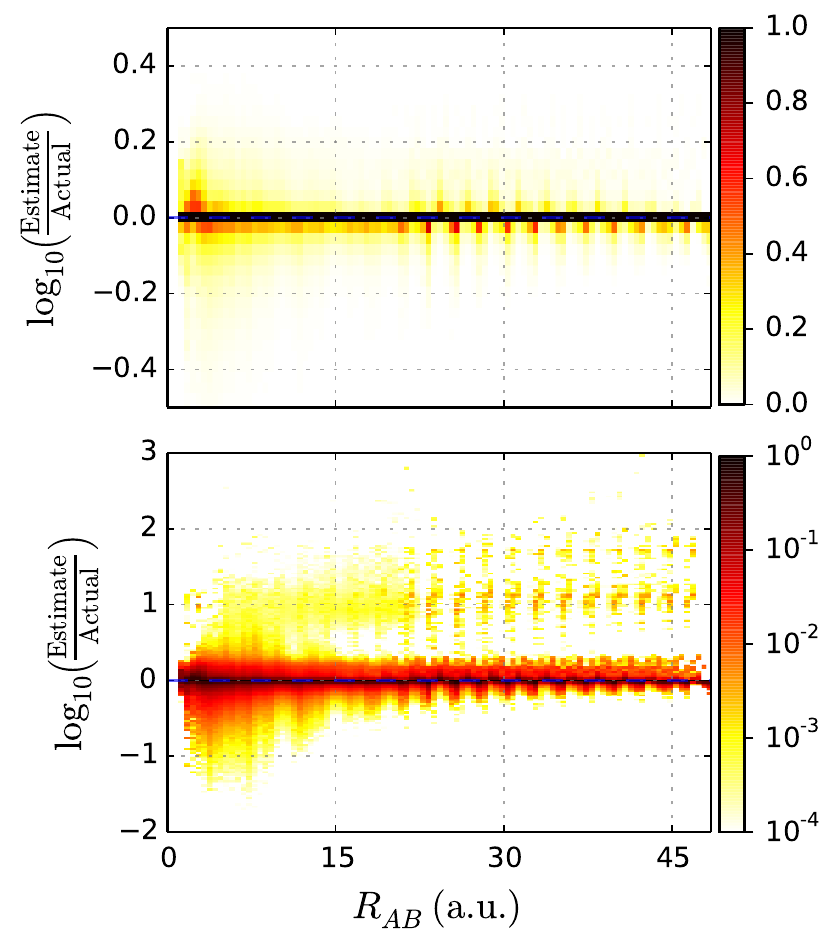}
  \caption{Heat map of the ratio $F = I_{\mathrm{estimate}}/I_{\mathrm{actual}}$ (well-separated integrals only) for our test set of molecules using the cc-pVTZ/cc-pV5Z/JK basis, with $\thws = 10^{-4}$ and $\thsq=10^{-1}$ using a linear (top) and logarithmic (bottom) color scale.  Ratios are normalized within each of 100 distance histogram bins so that the maximum in any given column is 1.0.  Note the difference in vertical scales between the two plots.}
  \label{fig:rvd_all}
\end{figure}

\begin{figure*}
  \includegraphics[width=\textwidth]{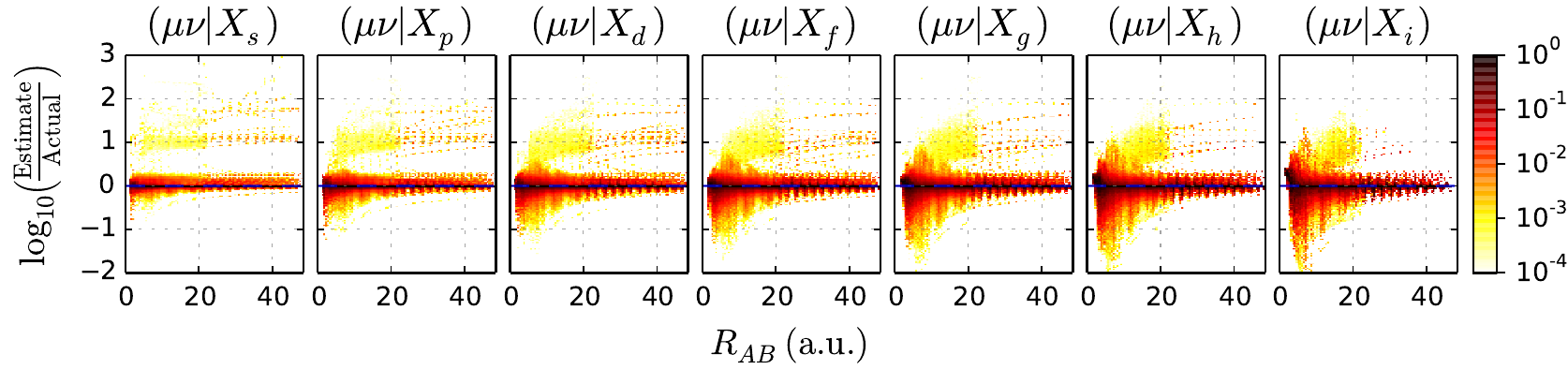}
  \caption{Data from Figure \ref{fig:rvd_all} split by angular momentum of the auxiliary basis function $\gk X$ using a logarithmic color scale.  See caption of Figure~\ref{fig:rvd_all} for details.}
  \label{fig:rvd}
\end{figure*}

\figref{rvd_all} shows a 2D histogram of the ratio $F$ plotted against bra--ket separation $R_{AB}$, with a linear color scale in the top plot and a logarithmic color scale on the bottom plot (note also the difference in the vertical scales). \figref{rvd} shows the same data split across ket angular momentum $\ell_X$.  In both figures, each distance bin is normalized individually to the range $[0, 1]$.  As with \figref{eva_all}, the linear color scale is relatively uninformative, except insofar as it shows that the vast majority of well-separated integrals are estimated almost exactly at all distances.  The logarithmic color scale reveals that most of the overestimates occur for shorter separations and taper off with increasing distance.  The same trend can particularly be seen for higher angular momentum in \figref{rvd}.  Again, this shows that the higher angular momentum contributions to the bra are more important for integrals with higher ket angular momentum.  At large bra-ket distances the estimator becomes more accurate due to the faster asymptotic decay of the contribution to the integral due to the higher angular momentum components to the bra.




\section{Conclusions}

We have introduced the \sqvl estimator for three-center ERIs. It is exact for some classes of integrals and is confirmed numerically to provide very tight estimates of the integrals for a wide variety of basis set types. The estimator incorporates the correct leading-order dependence on the bra-ket distance, thus significantly increasing the sparsity of the three-center ERI tensor in reduced-scaling electronic structure methods.  Computing the \sqvl estimate for a given shell triplet is relatively cheap (on the order of a couple dozen CPU clock ticks; a detailed analysis is dependent on the implementation details). Thus many algorithms will see modest performance gains simply by incorporating the estimate directly into the integral computation without any index reordering or extra bookkeeping.
More substantial gains are anticipated in algorithms that do not visit every shell triplet, even for the purposes of estimation.



Our basic tests performed outside of the context of any particular electronic structure method show that the norm of the most integrals in which the bra and ket are ``well-separated'' are estimated nearly perfectly, particularly for smaller basis sets that are more likely to be used for large molecule computations.  Our tests also show that the extent by which the integral norms are underestimated can be reduced readily without incurring explosion in computational cost. Tuning the adjustable parameters of the estimator ($\thws$ and $\thsq$) in the context of a given electronic structure method will be a function of the target precision and should be determined, as always in atomic basis electronic structure, by benchmarking. The use of the estimator for reduced-scaling construction of the Hartree-Fock exchange with the concentric atomic density fitting approximation\cite{Hollman:2013p064107} will be described in an upcoming manuscript. 


\section{Acknowledgements}

The work by DSH and EFV was supported by NSF grants CHE-1362655 and ACI-1047696, and Camille Dreyfus Teacher-Scholar Award.  The work by DSH and HFS was supported by NSF grant CHE-1361178.
This work used resources of the National Energy Research Scientific Computing Center, which is supported by the Office of Science of the U. S. Department of Energy under Contract No. DE-AC02-05CH11231.

\end{document}